\documentstyle[epsfig]{europhys}

\def\And{{\rm and\ }}

\def\order{\mathop{\rm O}\nolimits}

\def\stars{\bigskip\centerline{***}\medskip}

\newif\ifboo \boofalse

\def\Review#1{\boofalse{\it #1},}
\def\Name#1{{\sc #1},}
\def\Vol#1{\ifboo Vol. {\bf #1}\else{\bf #1}\fi}
\def\Year#1{\ifboo #1\else(#1)\fi}
\def\Book#1{\bootrue{\it #1},}
\def\Page#1{\ifboo {\rm p. #1}\else{\rm #1}\fi}

\begin{document}
\euro{0}{0}{0-0}{1998}
\Date{10 June 1998}
\shorttitle{A. LAMURA et al.: A LATTICE-BOLTZMANN MODEL OF TERNARY-FLUID MIXTURES}
\title{A Lattice Boltzmann Model of Ternary Fluid Mixtures}
\author{A. Lamura\inst{1}, G. Gonnella\inst{1}
  \And J. M. Yeomans\inst{2} }
\institute{
     \inst{1} Dipartimento di Fisica, Universit\'a di Bari, 
Istituto Nazionale di Fisica della Materia,\\
Unit\`a di Bari, and Istituto Nazionale di Fisica Nucleare, 
Sezione di Bari, \\
via Amendola 173, 70126 Bari, Italy \\
     \inst{2} Department of Physics, Theoretical Physics, 1 Keble Rd.,
Oxford OX1 3NP, England. }
\rec{}{}
\pacs{
\Pacs{47}{20.Ma}{Interfacial instability}
\Pacs{05}{70.Ln}{Nonequilibrium thermodynamics, irreversible processes}
\Pacs{83}{10.Lk}{Multiphase flows}
      }
\maketitle
\begin{abstract}
A lattice Boltzmann model is introduced which simulates
oil--water--surfactant mixtures. The model is based on a Ginzburg-Landau
free energy with two scalar order parameters. Diffusive and
hydrodynamic transport is included. Results are presented showing how
the surfactant diffuses to the oil--water interfaces thus lowering the
surface tension and leading to spontaneous emulsification. The rate of
emulsification depends on the viscosity of the ternary fluid.
\end{abstract}
\section{Introduction}
The addition of surfactant to a binary mixture of
oil and water can produce many different complex structures on
a mesoscopic length scale. 
The surfactant molecules move to the interface and lower the
oil--water interfacial tension. This can result in, for example,
lamellar, micellar, microemulsion or hexagonal arrangements of the oil
and water domains\cite{1,2}.

The equilibrium behaviour of such amphiphilic systems is well
understood. However the dynamics of the self-assembly of the mesoscale
phases and their rheology are less well investigated. This is a
difficult problem because of the interplay between several relevant
transport mechanisms, the diffusion of the constituent components and
their hydrodynamic flow. 
To date models of amphiphilic rheology which treat hydrodynamic effects
include time-dependent Ginzburg-Landau approaches\cite{3b,3c}, 
molecular dynamics\cite{4} and
a lattice gas cellular automaton scheme based on microscopic
interactions\cite{5b,5c}.

The aim of this Letter is to introduce an alternative numerical scheme that can
model the dynamics of amphiphilic systems in such a way that diffusive
and hydrodynamic mechanisms are included.
The numerical approach that we use is lattice Boltzmann simulations
which have emerged as a useful tool to study the dynamics of complex
fluids\cite{6}.  
We base our approach on that described by Orlandini {\em et.\
al.\ }\cite{7b, 7c}  where the
correct equilibrium of the fluid is imposed by
choosing an appropriate free energy and including it in such a way
that the fluid spontaneously reaches the equilibrium described by its
minimum.

Previous lattice Boltzmann models of amphiphilic systems have been
based on a single order parameter, that of the phase separating binary
fluid\cite{14,23}. The effect of the
amphiphilic molecules has been mimiced by varying the surface
tension in the input free energy. Although this approach proved
successful it
has the disadvantage
of not including the surfactant dynamics explicitly. 

We first give a description of the method and then 
present results showing how the surface
tension of the binary fluid interface is lowered by surfactant at a
rate which depends on the surfactant diffusion constant. As the 
surface tension 
becomes negative, this leads to the break-up of the interface and to
spontaneous emulsification\cite{25} to a lamellar phase. 

\section{The Lattice-Boltzmann Scheme}
We consider a Ginzburg-Landau model defined by the free energy functional
\cite{3b, 8b} which depends on two scalar order parameters
$\phi(\begin{bf}r\end{bf})$
and $\rho(\begin{bf} r\end{bf})$
\begin{eqnarray}
\begin{cal} F \end{cal}[\phi,\rho]&=& \int d \begin{bf} r \end{bf}
 \Big [\frac{a}{2}\phi^{2}+
                        \frac{b}{4}\phi^{4}+
                        \frac{\kappa}{2}(\nabla \phi)^{2}+
                        \frac{c}{2}(\nabla^{2}\phi)^{2}+  
     \frac{\alpha}{2}\rho^{2}+\frac{\lambda}{2}(\nabla\rho)^{2}+
                         \frac{\gamma}{2}(\nabla^{2}\rho)^{2} \nonumber\\
                                  & &+{}\beta_{1}\phi\rho^{2}+
                         \beta_{2}\phi^{2}(\nabla^{2}\rho)
                         +\beta_{3}\rho\phi(\nabla^{2}\phi)\Big ].
\label{fren}
\end{eqnarray}
$\phi(\bf r)$ and $\rho(\bf r)$ can be identified, respectively,
with the local density difference of oil and water and with the
difference of local surfactant concentration from its average
$\bar{\rho}$. $\bar{\rho}$ enters the model via the parameter
$\kappa$. $\kappa$ is
positive for small surfactant concentration.
As $\kappa$ decreases and
eventually becomes negative, $\bar{\rho}$ increases.

The thermodynamic variables that we will need are the chemical
potential difference between oil and water $\Delta \mu$, the chemical
potential $\Lambda$ of the surfactant and the pressure tensor
$P_{\alpha\beta}$. The chemical potentials  follow from the free 
energy as \cite{9}
\begin{equation}
\Delta \mu =  \frac{\delta{\cal F}}{\delta\phi}=a\phi+b\phi^{3}-\kappa\nabla^{2}\phi+c(\nabla^{2}\phi)^{2}+2\beta_{1}\rho\phi+2\beta_{2}\phi(\nabla^{2}\rho)+\beta_{3}\rho(\nabla^{2}\phi)+\beta_{3}\nabla^{2}(\rho\phi),
\label{mu}
\end{equation}
\begin{equation}
\Lambda  = \frac{\delta{\cal F}}{\delta\rho}=\alpha\rho-\lambda\nabla^{2}\rho+\gamma(\nabla^{2}\rho)^{2}+\beta_{1}\phi^{2}+\beta_{2}\nabla^{2}\phi^{2}+\beta_{3}\phi(\nabla^{2}\phi) .
\label{lam}
\end{equation}
The derivation of the pressure tensor is slightly more
complicated. Considering a linear combination of all symmetric tensors
having two or four gradient operators, we find that a suitable choice,
which allows the pressure tensor to obey the equilibrium condition
$\partial_{\alpha}P_{\alpha\beta}=0$ is
\begin{eqnarray}
P_{\alpha\beta}\!&=&\!\Big \{p_{L}+c \big [(\nabla^{2}\phi)^{2}+
\partial_{\sigma}\phi 
\partial_{\sigma}\nabla^{2}\phi \big ]+\gamma \big [
(\nabla^{2}\rho)^{2}+\gamma \partial_{\sigma}\rho
\partial_{\sigma}\nabla^{2}\rho \big ] 
+\beta_{2} \big[\partial_{\sigma}\phi^{2}\partial_{\sigma}\rho + \phi^{2}
\nabla^{2}\rho \big ] \nonumber \\
               \!& &\!+\beta_{3} \big [\partial_{\sigma}\rho \phi
\partial_{\sigma}\phi + \rho \phi \nabla^{2}\phi \big ] \Big \}
\delta_{\alpha\beta} + \kappa \partial_{\alpha}\phi \partial_{\beta}\phi
-c \big [ \partial_{\alpha}\phi \partial_{\beta} \nabla^{2}\phi+
\partial_{\beta}\phi \partial_{\alpha}\nabla^{2}\phi \big ] 
+ \lambda \partial_{\alpha}\rho \partial_{\beta}\rho \nonumber \\
               \!& &\!-\gamma \big [ \partial_{\alpha}\rho \partial_{\beta}
 \nabla^{2}\rho\!+\!\partial_{\beta}\rho \partial_{\alpha}\nabla^{2}\rho \big ] 
-\!\beta_{2} \big [
\partial_{\alpha}\phi^{2}\partial_{\beta}\rho\!+\!\partial_{\alpha}\rho
\partial_{\beta}\phi^{2}\big ]-\!\beta_{3} \big [\partial_{\alpha}\rho \phi
\partial_{\beta} \phi\!+\!\partial_{\alpha}\phi \partial_{\beta} \rho
\phi \big ] 
\label{pres}
\end{eqnarray}
where 
\begin{eqnarray}
p_{L}\!&\!=\!&\!\frac{a}{2}\phi^{2}\!+\frac{3}{4} b \phi^{4}\!-\kappa
\phi(\nabla^{2}\phi)
-\!\frac{\kappa}{2}(\nabla
\phi)^{2}\!+c\phi(\nabla^{2})^{2}\phi\!-\frac{c}{2}
(\nabla^{2}\phi)^{2}\!+\frac{\alpha}{2}\rho^{2}\!-\lambda
\rho(\nabla^{2}\rho)
-\!\frac{\lambda}{2}(\nabla \rho)^{2} \nonumber \\
     \!&\! \!&\!+\!\gamma \rho(\nabla^{2})^{2}\rho\!-\!\frac{\gamma}{2}
(\nabla^{2}\rho)^{2}\!+\!2\beta_{1}\rho
\phi^{2}\!+\!\beta_{2}\phi^{2}(\nabla^{2}\rho)\!+\!\beta_{2}\rho
(\nabla^{2}\phi^{2})
\!+\!\beta_{3}\rho \phi (\nabla^{2}\phi)\!+\!\beta_{3}\phi\nabla^{2}(\rho
\phi) 
\label{plon}
\end{eqnarray}
 
The lattice Boltzmann scheme is defined in terms of three 
distribution functions $f_{i}(\bf r)$, 
$g_{i}(\bf r)$ and $h_{i}(\bf r)$, each of which evolves 
during a time step $\Delta t$ according to
a single relaxation time Boltzmann equation \cite{10, 10b}
\begin{eqnarray}
f_{i}({\bf r}+{\bf e}_{i}\Delta t, t+\Delta t)-f_{i}({\bf r}, t)&=&
-\frac{1}{\tau}[f_{i}({\bf r}, t)-f_{i}^{0}({\bf r}, t)], \label{dist1}\\
g_{i}({\bf r}+{\bf e}_{i}\Delta t, t+\Delta t)-g_{i}({\bf r}, t)&=&
-\frac{1}{\tau_{\phi}}[g_{i}({\bf r}, t)-g_{i}^{0}({\bf r},
t)], \label{dist2} \\
h_{i}({\bf r}+{\bf e}_{i}\Delta t, t+\Delta t)-h_{i}({\bf r}, t)&=&
-\frac{1}{\tau_{\rho}}[h_{i}({\bf r}, t)-h_{i}^{0}({\bf r}, t)]
\label{dist3}
\end{eqnarray}
where $\tau$, $\tau_{\phi}$ and $\tau_{\rho}$ are independent
relaxation parameters and $\bf e_{\rm i}$ are the unit lattice
vectors. The distribution functions are related to the physical
variables by
\begin{equation}
n=\sum_{i}f_{i}, \hspace{0.5cm} n{\bf u}=\sum_{i}f_{i}{\bf e}_{i},\hspace{0.5cm}
\phi=\sum_{i}g_{i}, \hspace{0.5cm}\rho=\sum_{i}h_{i}
\label{phys}
\end{equation}
where n is the total density and $\bf u$ is the mean fluid velocity.
These quantities are locally conserved and, therefore, we require that
the equilibrium distribution functions $f_i^{0},g_i^0,h_i^0$  also 
fulfil Eqs.\ (\ref{phys}). The
higher moments of $f_{i}^{0}$, $g_{i}^{0}$ and $h_{i}^{0}$ are
defined by imposing the additional requirements
\begin{equation}
\sum_{i}f_{i}^{0}e_{i\alpha}e_{i\beta}=P_{\alpha\beta}+n u_{\alpha} u_{\beta} \;,
\; \sum_{i}g_{i}^{0}e_{i\alpha}=\phi u_{\alpha} \; , \; \sum_{i}g_{i}^{0}e_{i\alpha}
e_{i\beta}=\Gamma_{\phi}\Delta\mu\delta_{\alpha\beta}+\phi
u_{\alpha}u_{\beta} \;,
\end{equation}
\begin{equation}
\sum_{i}h_{i}^{0}e_{i\alpha}=\rho u_{\alpha} \; , \; \sum_{i}h_{i}^{0}e_{i\alpha}
e_{i\beta}=\Gamma_{\rho}\Lambda\delta_{\alpha\beta}+\rho
u_{\alpha}
u_{\beta}
\end{equation}
where $\Gamma_{\phi}$ and $\Gamma_{\rho}$ are mobilities. The
equilibrium distribution functions are defined as usual in terms of an
expansion in the velocity $\bf u$\cite{7b,7c}. We use a 9-velocity model on a
square lattice to obtain the results presented here.

These
definitions lead to the continuum equations which follow from
expanding Eqs.\ (\ref{dist1}), (\ref{dist2}) and (\ref{dist3}) to $\order (\Delta t^{2})$\cite{7c}
\begin{equation}
\partial_{t}n+\partial_{\alpha} (n u_{\alpha})=0 \; , \;
\partial_{t}(n u_{\alpha})+\partial_{\beta} (n
u_{\alpha}u_{\beta})=-\partial_{\beta}P_{\alpha \beta}+\nu
\nabla^{2}(n
u_{\alpha})+\partial_{\alpha} \big [\lambda(n)\partial_{\gamma}(n
u_{\gamma}) \big ] \; ,
\label{nast}
\end{equation}
\begin{equation}
\partial_{t}\phi+\partial_{\alpha}(\phi
u_{\alpha})=\Gamma_{\phi}\Theta_{\phi}\nabla^{2}\Delta
\mu-\Theta_{\phi}\partial_{\alpha}\left(\frac{\phi}{n}
\partial_{\beta}P_{\alpha
\beta}\right),
\label{conv1}
\end{equation}
\begin{equation}
\partial_{t}\rho+\partial_{\alpha}(\rho
u_{\alpha})=\Gamma_{\rho}\Theta_{\rho}
\nabla^{2}\Lambda-\Theta_{\rho}\partial_{\alpha}\left(\frac{\rho}{n}
\partial_{\beta}P_{\alpha
\beta}\right),
\label{conv2}
\end{equation}
where 
\begin{equation}
\nu=\frac{(2\tau-1)}{6}(\Delta t),\ \lambda(n)=(\tau-\frac{1}{2})\Delta
t(\frac{1}{2}-\frac{dp_{0}}{dn}),\ \Theta_{\phi}=\Delta t(\tau_{\phi}-\frac{1}{2}),\
\Theta_{\rho}=\Delta t(\tau_{\rho}-\frac{1}{2}).
\label{param}
\end{equation}
\section{Dynamical Behaviour}
Unless otherwise stated the simulations were run with $a=-1$, 
$b=c=1$, $\kappa=0.1$,
$\alpha=\lambda=\gamma=1$, $\beta_{1}=0$, $\beta_{2}=-0.2$, $\beta_{3}=0.4$,
$\tau=100$, $\tau_{\phi}=\tau_{\rho}=(1+1/\sqrt{3})/2$ \cite{7b} and with units in which
$\Delta t=1$.

Above the critical temperature ($a > 0$) the equilibrium
configuration is a mixture of the three components, with constant $n$,
 $\phi=0$ and $\rho=0$. For small variations in $\phi$, Eq.\ 
(\ref{conv1}) can be
linearised about $\phi=0$ resulting in a convection--diffusion equation with
diffusion constant
$D_{\phi}=a\Gamma_{\phi}\Theta_{\phi}$. Similarly 
for small variations in $\rho$, Eq.\ (\ref{conv2}) can be linearised
resulting in a diffusion equation with
$D_{\rho}=\alpha\Gamma_{\rho}\Theta_{\rho}$. Following \cite{7b}, we
tested
these predictions by measuring $D_{\phi}$ and  $D_{\rho}$ as functions of
$\Gamma_{\phi}\Theta_{\phi}$ and $\Gamma_{\rho}\Theta_{\rho}$
respectively by monitoring the decay of a sinusoidal perturbation. 
The measured values are shown in Fig.\ 1
for different values of $a$ and $\alpha$. Agreement with the predicted
values is very good.

\begin{figure}
\begin{center}
   \includegraphics*[width=7.0 cm,height=6 cm]{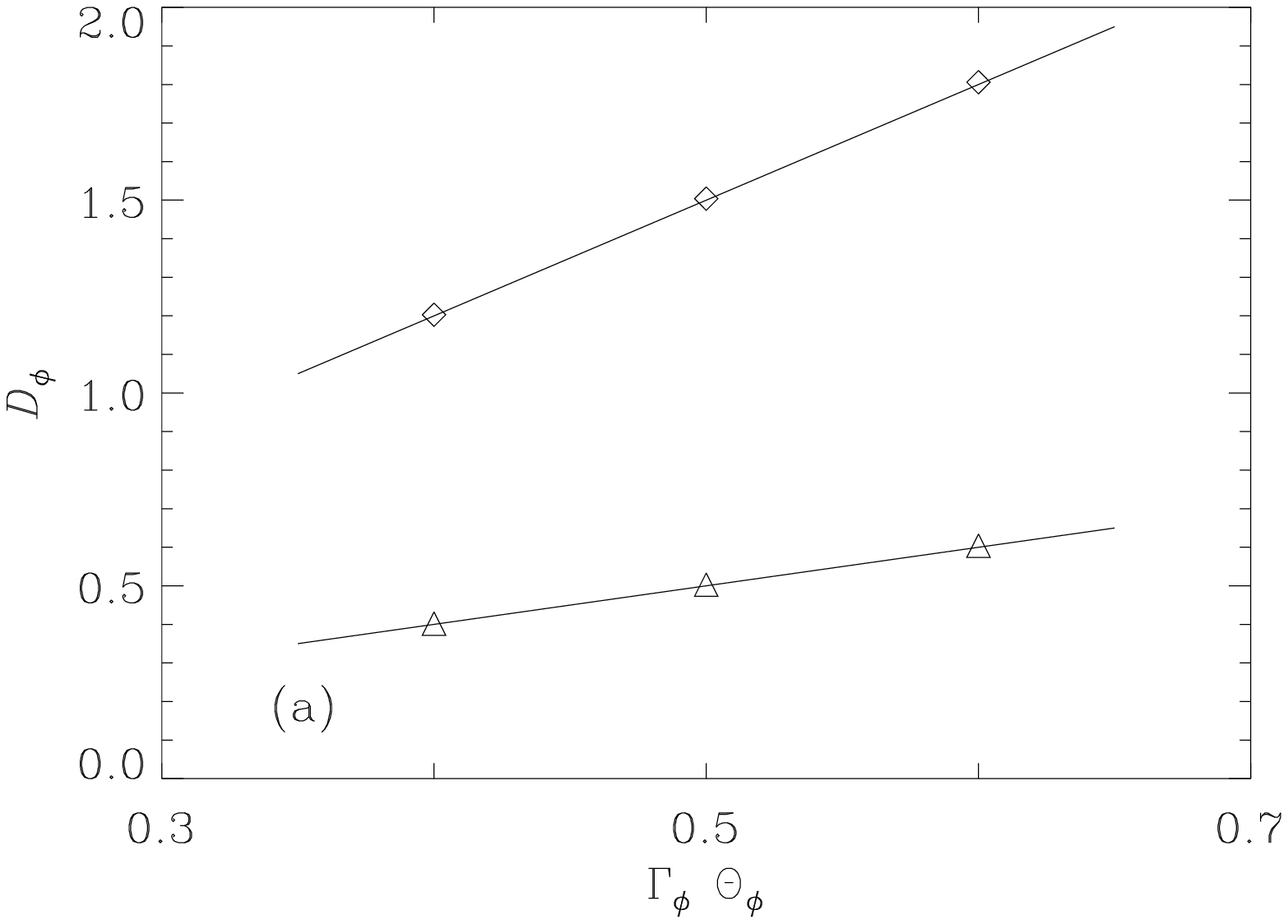}
   \includegraphics*[width=7.0 cm,height=6 cm]{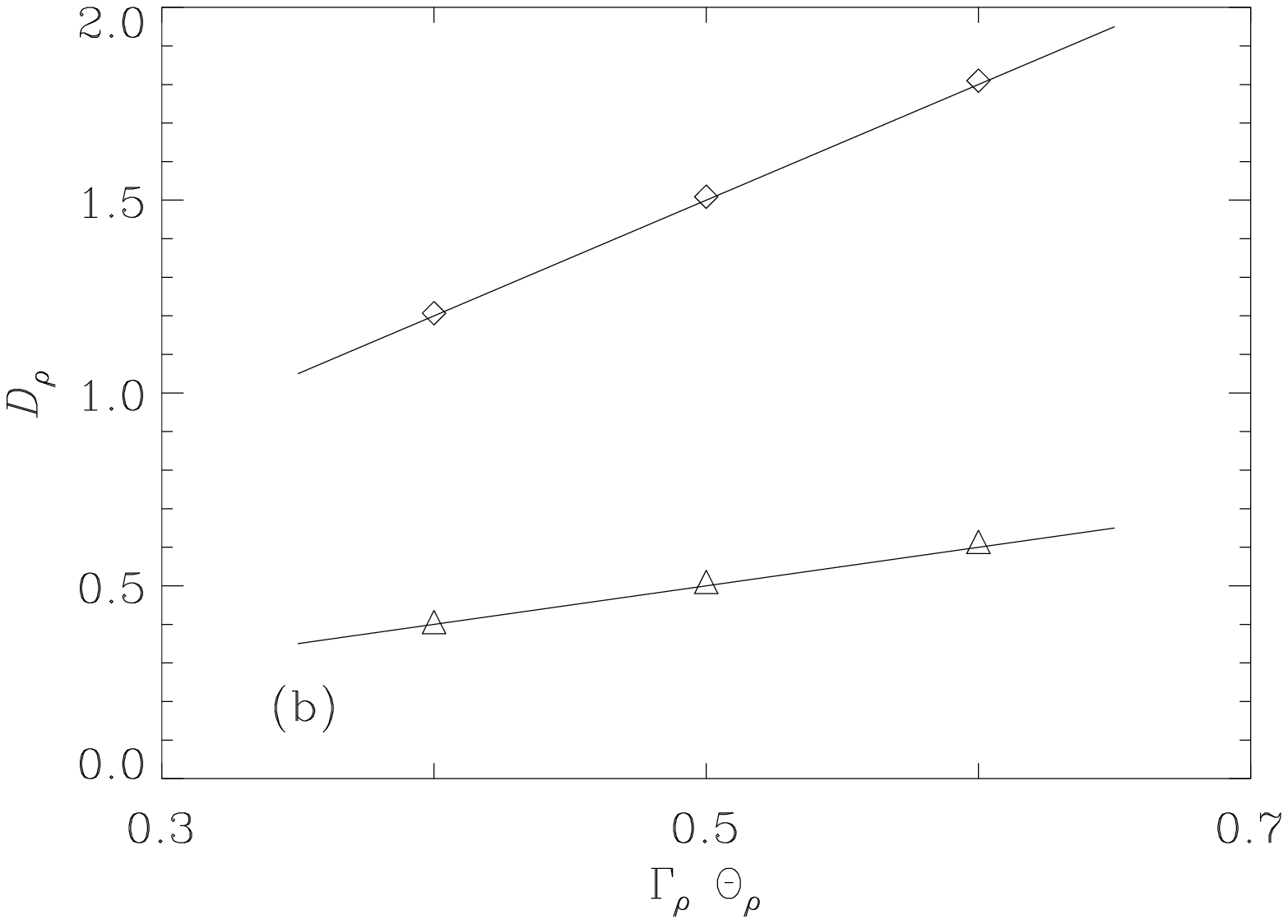}\\*
\end{center}
\caption{Diffusion constant $D$ as a function of mobility
$\Gamma \Theta$ for temperatures above critical for (a) the oil--water
density difference and (b) the surfactant density.
 The full lines are analytic results which follow
from linearising Eqs.\ (\ref{conv1}) and (\ref{conv2}). Data points were
 obtained by
following the decay of a small sinusoidal perturbation.}
\begin{center}
   \includegraphics*[width=7.0 cm,height=6 cm]{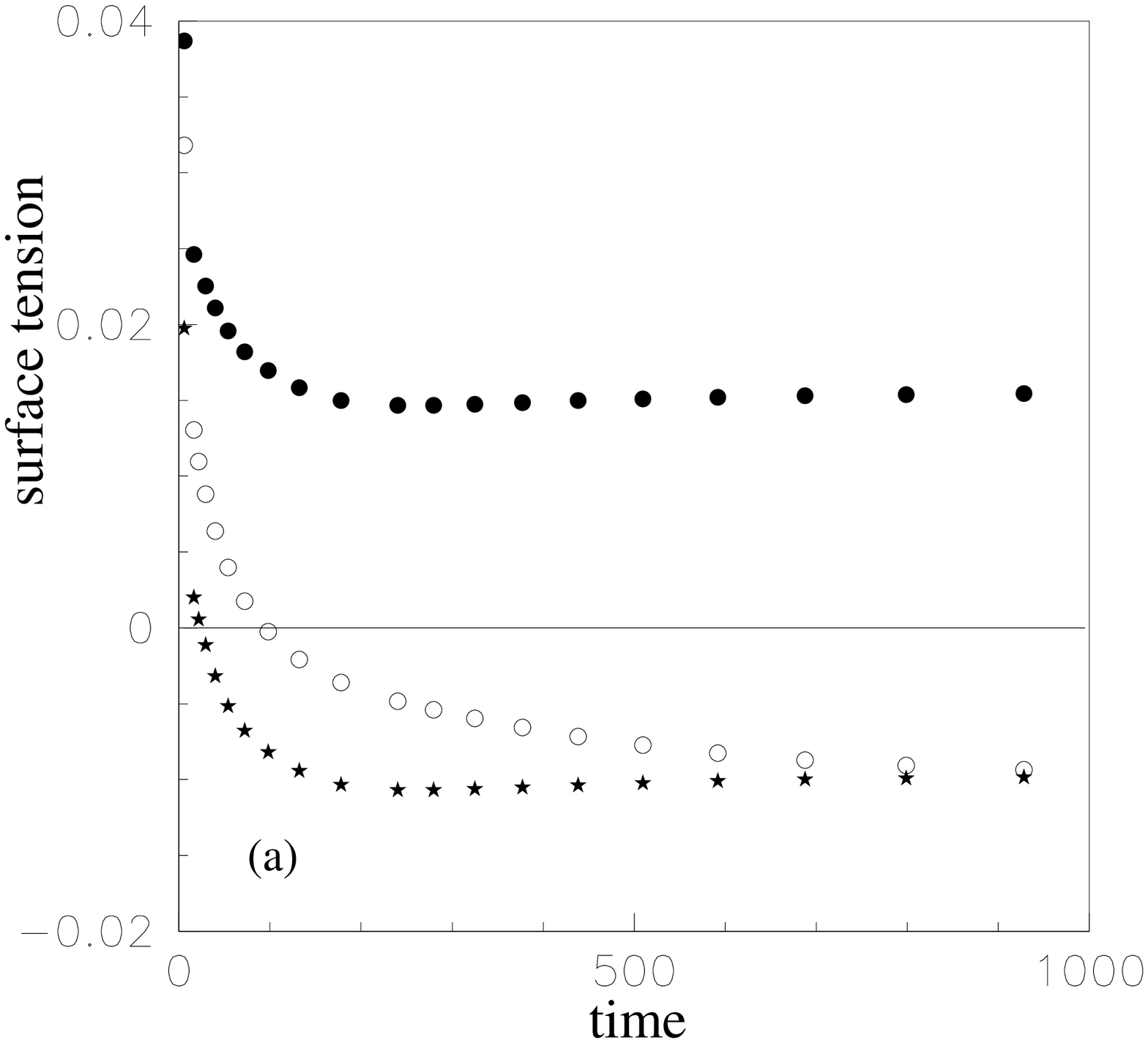}
   \includegraphics*[width=7.0 cm,height=6 cm]{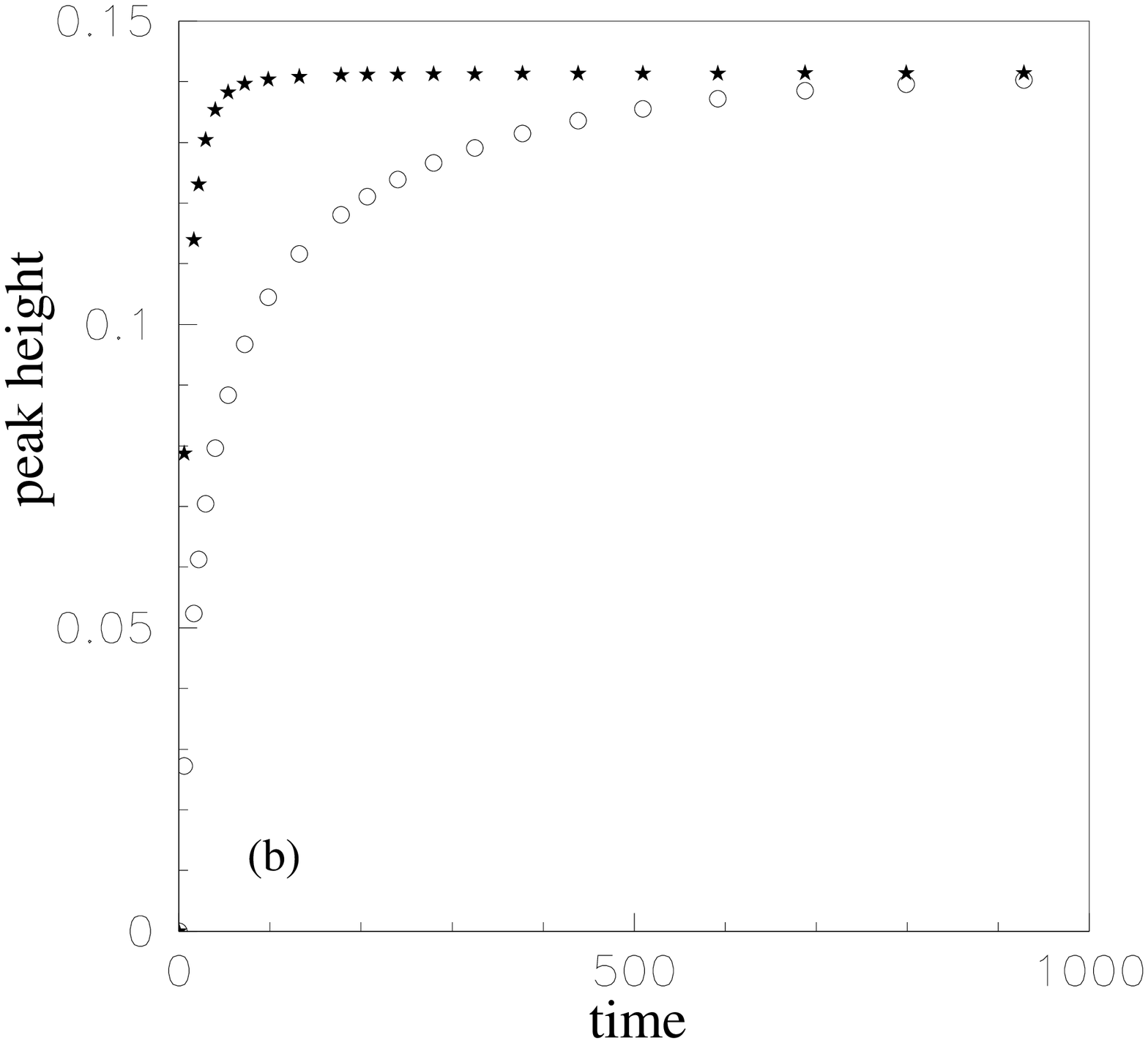}\\*
\end{center}
\caption{Variation of (a) the surface tension of a flat interface, (b)
the maximum surfactant density at the interface with time for
different values of the surfactant mobility: 
$\bullet \; \Gamma_{\rho}\Theta_{\rho}=0$, $\circ \; 
\Gamma_{\rho}\Theta_{\rho}=0.1$,
$\star \; \Gamma_{\rho}\Theta_{\rho}=2.5$. The parameter $\kappa$ is fixed
to the value $-1.15$ and $\Gamma_{\phi}\Theta_{\phi}=0.1$.}
\end{figure}

Below the critical temperature ($a < 0$) the binary mixture phase separates
into two distinct phases, symmetric about $\phi=0$. Oil--water interfaces
are now formed in the system. We next consider the
diffusion of the surfactant towards these interfaces.
A system set-up to contain an oil--water interface
that was initially a hyperbolic tangent was initialised with $\rho=0$ and  
and then allowed to evolve.
Figure\ 2a shows  the
surface tension plotted as a function of time for three different
values of the surfactant mobility. For zero mobility the interface
relaxes to its equilibrium shape thus causing a decrease in the
surface tension. However, no surfactant diffuses to the interface and
the surface tension remains positive. For finite
mobility however the surfactant can diffuse to the interface and the final
surface tension is negative. 
\begin{figure}
\begin{center}
   \includegraphics*[width=11.5 cm, height=14.5 cm]{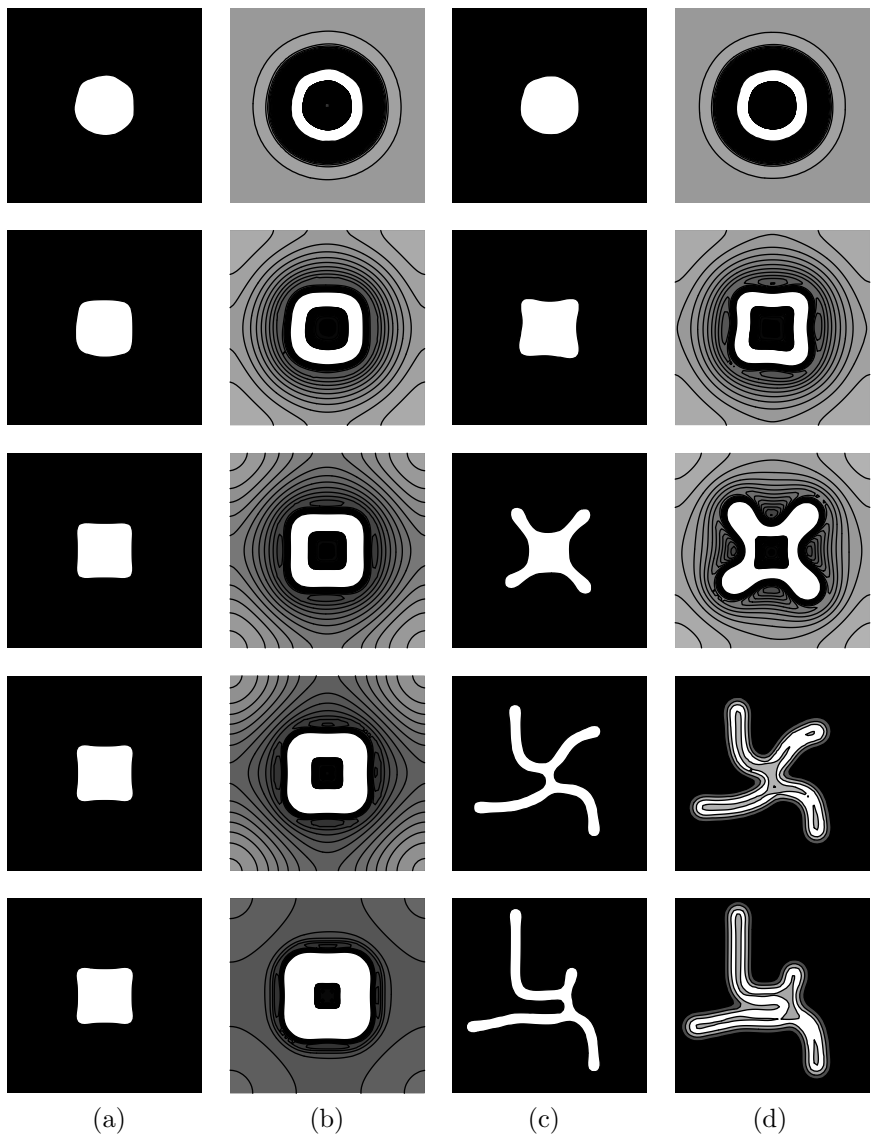}\\*
(a)\hspace{0.17\textwidth}(b)\hspace{0.17\textwidth}(c)
\hspace{0.17\textwidth}(d)
\end{center}
\caption{Spontaneous emulsification of a droplet as surfactant
diffuses to the interface. Results are given for the evolution with
time (from top, t=$46$, $591$, $1252$, $2652$, $5615$) of (a) the
oil--water density difference $\phi$ for high viscosity; (b) the 
surfactant density $\rho$ (grey-scaling from black $\Rightarrow$ white
corresponds to minimum $\rho$ $\Rightarrow$ maximum $\rho$)
for high viscosity; (c) $\phi$ for low
viscosity; (d) $\rho$ for low viscosity.
$\kappa$ =
$-1.15$ and $\Gamma_{\phi}\Theta_{\phi}=\Gamma_{\rho}\Theta_{\rho}=0.1$.}
\label{fig3}
\end{figure}
The rate at which the final value is
achieved clearly depends on the value of the diffusion constant. Similar
considerations apply to the increase in the peak value of the
surfactant density $\rho$ at the interface as a function of time
(Fig.\ 2b).

Once the surface tension has become negative the interface should
become unstable\cite{25}. 
Figure\ 3 shows the evolution with time of a drop of oil
in water which is initially approximately circular.
(Small randomness in the radius is necessary or the drop remains metastable.)
The evolution of both the $\phi$ and
$\rho$ fields are shown. Small perturbations on the surface of
the drop grow to form lamellae of a width appropriate to that
minimising the free energy. The budding of the
drop initially reflects the symmetry of the underlying lattice. We
believe this to be a consequence of the sharpness of the interface
relative to the lattice spacing.

Figure\ 3 compares the spontaneous emulsification at high ($\tau=100$)
 and low ($\tau=0.585$)
viscosities. The change in viscosity makes a striking
difference to the speed at which the lamellae form. 

\section{Summary} 
We have described a lattice Boltzmann model for ternary mixtures, such
as oil--water--surfactant systems. The model includes hydrodynamic and
diffusive modes and controls the fluid equilibrium via a chosen input
free energy. The viscosity and diffusivities of the fluid phases can
be changed thus allowing a study of their effects on the self-assembly
and rheology of the amphiphilic fluid. In particular we presented
results for the effect of surface diffusion on the rate of change of
the interfacial tension and for the spontaneous emulsification which
results as the tension becomes negative. 

The model is not trivial and much work remains to be done to explore
its physical and numerical properties within a wide dynamic and
static parameter space. A well-defined equilibrium and
the ability to impose rather than measure transport coefficients are
particularly useful in this task. Extensions to three dimensions and a
closer comparison to experimental results are important endeavours. 

\stars

We thank Enzo Orlandini and Peter Coveney for helpful discussions.


\vskip-12pt
\begin{thebibliography}{99}
%
\bibitem{1}
\Name{Gelbart W. M., Ben-Shaul A. \And Roux D.} editors,
\Book{Micelles, Membranes, Microemulsions and Monolayers} 
Springer-Verlag, Berlin
\Year{1994}.
%
\bibitem{2}
\Name{Gompper G. \And Schick M.} in \Book{Phase Transition and
Critical Phenomena} edited by \Name{C. Domb \And J. L. Lebowitz}
\Vol{16} (Academic Press, London) \Year {1994} p\Page{1-176}.
%
\bibitem{3b}
\Name{Hennes M. \And Gompper G.} \Review{Phys. Rev. E} \Vol{54}
\Year{1996} \Page{3811}.
%
\bibitem{3c}
\Name{P\"{a}tzold G. \And Dawson K. A.}
 \Review{Phys. Rev. E}
 \Vol{52}
\Year{1995} \Page{6908}.
%
\bibitem{4}
\Name{Laradji M., Mouritsen O. G., Toxvaerd S. \And Zuckerman M.J.} 
\Review{Phys. Rev. E} \Vol{50}
 \Year{1994} \Page{1243}.
%
%
\bibitem{5b}
\Name{Emerton A. N., Coveney P. V. \And Boghosian B. M.}
 \Review{Phys. Rev. E} \Vol{55} \Year{1997} \Page{708}.
\bibitem{5c}
\Name{Weig F. W. J., Coveney P. V. \And Boghosian B. M} \Review{Phys. Rev. E}
\Vol{56} \Year{1997} \Page{6877}.
%
%
\bibitem{6}
\Name{Chen S. \And Doolen G. D.}
\Review{Ann. Rev. Fluid. Mech.} \Vol{30} \Year{1998} \Page{329}
%
\bibitem{7b}
\Name{Orlandini E., Swift M. R. \And
Yeomans J. M.} \Review{Europhys. Lett.} \Vol{32} \Year{1995} \Page{463}.
%
\bibitem{7c}
\Name{Swift M. R., Orlandini E., Osborn W. R. \And Yeomans J. M.}
 \Review{Phys. Rev. E} \Vol{54} \Year{1996} \Page{5041}.
%
\bibitem{14}
\Name{Gonnella G., Orlandini E. \And Yeomans J. M.} \Review{Phys. Rev. Lett.}
 \Vol{78}
 \Year{1997}  
 \Page{1695}.
%
\bibitem{23}
\Name{Theissen O., Gompper G. \And Kroll D.M.}
\Review{Europhys. Lett.}
\Vol{42}
\Year{1998}
\Page{419}.
%
\bibitem{25}
\Name{Granek R., Ball R. C. and Cates M. E.}
\Review{J. Phys. II France}
\Vol{3}
 \Year{1993} \Page{829}.
%
\bibitem{8b}
\Name{Gompper G \And Schick M.} \Review{Phys. Rev. E}
 \Vol{49} \Year{1994} \Page{1478}.
%
\bibitem{9}
\Name{Rowlinson J. S. \And Widom B.} \Book{Molecular Theory of Capillarity} 
Oxford
\Year{1982}.
%
\bibitem{10}
\Name{Bhatnagar P. L., Gross E. P. \And Krook M.} \Review{Phys. Rev.}
 \Vol{94} \Year{1954}  
 \Page{511}.
%
\bibitem{10b}
\Name{Chen H., Chen S. \And Matthaeus W. H.} \Review{Phys. Rev. A} \Vol{45}
 \Year{1992}  
 \Page{R5339}.
\end{thebibliography}
\end{document}